\documentclass[useAMS,usegraphicx,usenatbib]{mn2e}
\usepackage{amsfonts,amssymb,amsmath}
\usepackage{xcolor}
\usepackage{hyperref}
\usepackage{graphicx}

\title[Low X-ray Luminosity Galaxy Clusters. III: Weak Lensing Mass Determination at 0.18 $<$ z $<$ 0.70.]{Low X-ray Luminosity Galaxy Clusters. III: Weak Lensing Mass Determination at 0.18 $<$ z $<$ 0.70.}
\author[Gonzalez et al.]{Elizabeth Johana Gonzalez$^{1,2}$\thanks{E-mail:
elizabethjgonzalez@oac.unc.edu.ar}, 
Gael Fo\"{e}x $^{3}$, 
Jos\'e Luis Nilo Castell\'on$^{4}$,
  \newauthor 
Mariano J. Dom\'{\i}nguez Romero$^{1,5}$,
Mar\'ia Victoria Alonso$^{1,5}$,
 \newauthor 
Diego Garc\'{\i}a Lambas$^{1,5}$,
Osvaldo Moreschi$^{2,6}$,
Emanuel Gallo$^{2,6}$\\
$^{1}$ Instituto de Astronom\'{\i}a Te\'orica y Experimental, (IATE-CONICET),
 Laprida 854, X5000BGR, C\'ordoba, Argentina.\\
$^{2}$ Facultad de Matem\'atica, Astronom\'ia y F\'isica, FAMAF, Universidad Nacional de C\'ordoba, X5000BGR, C\'ordoba, Argentina.\\
$^{3}$ Instituto de F\'isica y Astronom\'ia, Universidad de Valpara\'iso, Avda. Gran Bretaña 1111, Valpara\'iso, Chile\\
$^{4}$ Departamento de F\'isica y Astronom\'ia, Facultad de Ciencias, Universidad de La Serena. Avenida Juan Cisternas 1200, La Serena, Chile.\\
$^{5}$ Observatorio Astron\'omico de C\'ordoba, Universidad Nacional de C\'ordoba, Laprida 854, X5000BGR, C\'ordoba, Argentina.\\
$^{6}$ Instituto de F\'isica Enrique Gaviola (IFEG-CONICET), Medina Allende s/n, Ciudad Universitaria X5016LAE C\'ordoba, Argentina. }

\begin{document}


\pagerange{\pageref{firstpage}--\pageref{lastpage}} \pubyear{2015}

\maketitle

\label{firstpage}

\begin{abstract}

This is the third of a series of papers of low X-ray luminosity
galaxy clusters. In this work we present the weak lensing analysis of eight clusters,
based on observations obtained with the Gemini Multi-Object Spectrograph in the $g'$, $r'$ and $i'$ passbands. For this purpose, we have developed a pipeline for the lensing analysis of
ground-based images and we have performed tests applied to simulated data. \\
We have determined the masses of seven galaxy clusters, six of them measured for the first time. For the
four clusters with availably spectroscopic data, we find a general agreement between the
velocity dispersions obtained via weak lensing assuming a Singular Isothermal Sphere profile, and those
obtained from the redshift distribution of member galaxies. The correlation between our weak lensing mass determinations and the X-ray luminosities are suitably fitted by other observations of the $M-L_{X}$ relation and models. 

\end{abstract}


\section{Introduction}

Clusters of galaxies  are  the  most  massive vi\-ria\-li\-zed  structures  in  the  Universe. Hence, they are excellent laboratories to study the physics of baryonic and dark matter at large scales in bound objects \citep{Voit05,Pratt09,Arnaud10,Giodini13}.  Numerical si\-mu\-la\-tions show that massive clusters are formed from the mer\-ging of smaller structures in the hierarchical structure formation \citep[see review,][]{Kravtsov12}. Therefore, the study of low X-ray galaxy clusters could shed light on the assembly processes and environmental e\-ffects on their galaxy population, since these systems are likely to be e\-vol\-ving by sub-structure interations and accretion. In these systems, velocity dispersions are lower than in massive cluster ($\lesssim 800$\,km\,s$^{-1}$), favoring the interactions and mergers between the galaxy members. Thus, morphological transformations are more frequent in these clusters. Also, low mass clusters are more common than rich clusters due to the steepness of the cluster mass function. However, at the same time these systems are fainter and cooler, which makes them more difficult to detect and distinguish from background. Hence, these clusters have not been extensively studied compared to massive, luminous X-ray systems.

The evolution of galaxy clusters has been probed to be determined by cosmological parameters. In particular, the cluster mass function provides observational constraints to cosmology, given its sensitivity on the cosmological pa\-ra\-me\-ters \citep[e.g.,][]{Mandelbaum07,Rozo09,Vikhlinin09,Allen11,Planck14}. The main limitation in the use of this mass function is the practical determination of the masses. Weak and strong gavitational  lensing  probe  the  projected  mass  distribution  of  clusters,  with strong lensing confined to the central regions of clusters, whereas weak lensing can yield mass measurements for larger radii. Mass estimations from gravitational lensing is affected by substructure, triaxiality, large-scale structure and the po\-ssi\-ble presence of multiple haloes along the line-of-sight
 \citep{Oguri05,Sereno07,Corless09,Meneghetti10,Sereno10,Sereno11,Giocoli12,Sereno12,Spinelli12}. However, other me\-thods such as the caustic technique employing spec\-tros\-co\-pic measurements of galaxies velocity \citep{Rines06}, might be expensive in telescope time. Besides, radial mass distribution of clusters could be determined u\-sing X-ray surface brightness under the assumption of hydrostatic e\-qui\-li\-brium \citep{LaRoque06,Donahue14}. Nevertheless, deviations from the equilibrium could highly affect the estimations. Therefore, gravitational len\-sing is an excellent and a fairly clean technique for mass cluster determinations. 
 
Galaxy clusters and groups are expected to fo\-llow simple relations linking the total mass with another physical quantities \citep{Kaiser86}. Given the difficulties of determining the mass of these systems, the study of these relations are important since they are suitable to convert simple observables into mass estimates. In particular, the X-ray luminosity of groups and clusters can be considered a good tracer of halo masses with approximately 20$\%$ scatter in the M-L$_{X}$ relation \citep{Stanek06, Maughan07, Pratt09, Rozo08,Rykoff08,Vikhlinin09b}. The main advantage in its use is that X-ray luminosity can be accurately measured at high redshifts, re\-qui\-ring only previous cluster detection and redshift information. Weak lensing provides a suitable technique to study the M-L$_{X}$ relation and it has been recently a\-pplied in several works \citep{Bardeau07,Hoekstra07,Rykoff08,Leauthaud10,Okabe10}. In this sense, three studies s\-pa\-nning from low \mbox{X-ray} luminosity clusters to groups \citep{Rykoff08,Leauthaud10,Kettula14} show a single relation with a well defined slope \citep{Foex12}, in agreement with those of massive clusters.
 
This work is the third in a series of papers aimed to understand the processes involved in the formation and evolution of low X-ray luminosity galaxy clusters at in\-ter\-me\-dia\-te redshifts. The first paper of the series \citep[][hereafter Paper I]{PaperI} contains the main goals, sample selection, and details of observations and data reduction for both, photometry and spectroscopy. The second paper \citep[][hereafter Paper II]{PaperII}, presents photometric properties of seven low X-ray luminosity observed with Gemini telescopes. As the redshift increases, an increment of blue galaxies and a decline in the fraction of lenticulars is observed, while the early-type fraction remains almost constant. These results are in agreement with those for high mass clusters. At lower redshifts, the presence of a well-defined cluster red sequence extending by more than 4 magnitudes showed that these intermediate mass clusters had reached a relaxed stage. 

In this oportunity we present the weak lensing analysis of eight galaxy clusters of the low X-ray luminosity sample. The paper is organizated as follow: In Sec.\,\ref{sec:sample}, we describe the sample of clusters, and the acquisition and reduction of the images. In Sec.\,\ref{sec:method} we give the details of the weak lensing analysis for the mass determination. In Sec.\,\ref{sec:results} we present and discuss the estimated mass, and compared them with X-ray luminosity. Finally, in Sec.\,\ref{sec:conclusions} we summarise the main results of this work. We adopt when necessary a standard cosmological model $H_{0}$\,=\,70\,km\,s$^{-1}$\,Mpc$^{-1}$, $ \Omega_{m} $\,=\,0.3, and $ \Omega_{\Lambda} $\,=\,0.7.

\section{GALAXY CLUSTERS, OBSERVATIONS AND DATA REDUCTION}

\begin{table*}
\caption{Low X-ray luminosity Galaxy Cluster sample}
\label{table1}
\begin{tabular}{crrrrclccc}
\hline
[VMF98] & $\alpha$ &   $\delta$      &       $L_X$   &  $L_X$     &z & Program & g$^{\prime}$ & r$^{\prime}$ & i$^{\prime}$\\
           &  &         &        [0.5--2.0] keV  & [0.1--2.4] keV    & &  &  &  & \\
           Id.    &(J2000)          &         (J2000)    &   ($h^{-2}_{70}$10$^{43}$ cgs)& ($h^{-2}_{70}$10$^{43}$ cgs)  & & Id. & &              &             \\
         \hline
          001    & 00 30 33.2  & +26 18 19  & 26.1 & 30.7    & 0.500 & GN-2010B-Q-73    & --        &15$\times$300 & 15$\times$150  \\
          022    & 02 06 23.4  & +15 11 16  & 3.6 & 3.8 & 0.248 &  GN-2003B-Q-10    & --        & 4$\times$300 & 4$\times$150 \\
          093    & 10 53 18.4  & +57 20 47  & 1.4 &1.6 & 0.340 &  GN-2011A-Q-75    & --         & 5$\times$600 & 4$\times$150 \\
          097    & 11 17 26.1  & +07 43 35 & 6.4 & 7.7 & 0.477 & GS-2003A-SV-206    & 12$\times$600 & 7$\times$900 & -- \\
          102    & 11 24 13.9  & -17 00 11 & 8.1 & 9.3 &0.407 &  GS-2003A-SV-206    & -- & 5$\times$600 & -- \\         
          119    & 12 21 24.5  & +49 18 13  & 42.7 &53.6   & 0.700 & GN-2011A-Q-75    & --        & 7$\times$190 & 4$\times$120 \\
          124    & 12 52 05.4  & -29 20 46 & 3.4   & 3.4 &0.188 & GS-2003A-SV-206  & 5$\times$300  & 5$\times$600 & -- \\
          148    & 13 42 49.1 & +40 28 11  & 16.2& 21.4 &0.699 & GN-2011A-Q-75      & --           & 7$\times$190 & 5$\times$120 \\
 \hline
 \end{tabular}
 \medskip
 \begin{flushleft}
\textbf{Notes.} Columns: (1), the cluster identification; (2) and (3), the equatorial 
coordinates of the X-ray centre; (4), the X-ray
luminosity in the [0.5 - 2.0] keV energy band obtained from \citet{Vikhlinin98}; (5), shows the X-ray luminosity in the [0.1 - 2.4] keV energy band calculated using $L_{500}$ from the MCXC catalogue \citep[\textit{Meta-Catalogue of X-ray Detected Clusters of Galaxies,}][]{Piffaretti11}; (6), the mean redshift for each cluster from \citet{Mullis03}; (7), the Gemini Program identification;
(8), (9) and (10), the number of exposures and individual exposure time in 
seconds for each passband.

\end{flushleft}
\end{table*}

\subsection{Sample description}
\label{sec:sample}

The studied sample of low X-ray luminosities was selected from the catalogue of extended X-ray sources by \citet{Mullis03}. This catalogue is a revised version of the 223 galaxy clusters serendipitously detected in the ROSAT PSPC pointed observations by \citet{Vikhlinin98}.  Our galaxy cluster sample comprises a random selection of 19 systems from the total sample of 140 galaxy clusters with X-ray luminosities in the [0.5--2.0]\,keV energy band (rest frame), close to the detection limit of the ROSAT PSPC survey ranging from $10^{42}$ to $\sim 50 \times 10^{43}$ erg s$^{-1}$. The redshift range of our selection is 0.16 to 0.70 and a full description of the project and sample can be found in Paper\,I.

The galaxy clusters subsample studied in this work is mainly based on the clusters optically analized in Paper II: 7 galaxy clusters  with X-ray luminosity ranging from  1.4 to 26.1 $\times$10$^{43}$  ergs$^{-1}$ in the [0.5--2.0]\,keV energy band,  and redshifts between 0.185 to 0.7.  We add to this sample with observed colours, the galaxy cluster [VMF98]102 located at $ z\sim 0.401$, observed only in $r'$ passband. In Table\,\ref{table1} we summarize the main characteristics of the clusters. The mean X-ray luminosity in [0.5--2.0] keV band is 13.4 $\times$ 10$^{43}$ erg s$^{-1}$ , an intermediate/low luminosity when compared to $\sim$10$^{42}$  erg s$^{-1}$ for groups with extended X-ray emission or the larger values than 5 $\sim$ 10$^{44}$ erg s$^{-1}$ of rich clusters. $L_{X}$ in [0.1--2.4]\,keV band are used for further analysis and comparison with other workes (see Section\,\ref{lmrelation})

\subsection{Observations}

Photometric observations for the eight galaxy clusters were obtained with Gemini North (GN) and South (GS) telescopes, during the system verification process (SVP) and specific programs with Argentinian time allocation. Seven clusters were observed using the Gemini Multi-Object Spectrograph \citep{Hook04} in the image mode, in the $r'$ and $g'$ or $i'$ passbands with an array of three EEV CCDs of $2048\times4608$ pixels and only one ([VMF98]102) in the $r'$ passband. Using a $2\times2$ binning, the pixel scale is 0.1454 arcsec per pixel which corresponds to a FOV (\textit{Field of View}) of $5.5\times5.5$ arcmin$^{2}$ in the sky.

All images were observed under excellent photometric conditions, with mean seeing values of 0.75, 0.66 and 0.74 arcsec in the $g'$ , $r'$ and $i'$ filters, respectively. Some observations were made under exceptional weather conditions, such as those made to the galaxy cluster [VMF98]001, with a median seeing of about 0.485 in the $r'$ image.  Further details about these observations are given in Paper\,II.  Columns 6 to 9 in Table\,\ref{table1} show a summary of the photometric observations.  

All observations were processed with the Gemini IRAF package v1.4 inside IRAF\footnote{IRAF is distributed by the National Optical Astronomy Observatories, which are operated by the Association of Universities for Research in Astronomy, Inc., under cooperative agreement with the National Science Foundation.}  \citep{Tody93} . The images were bias/overscan-subtracted, trimmed and flat-fielded. The final processed images were registered to a common pixel position and then combined.

\section{WEAK LENSING ANALYSIS}
\label{sec:method}
We developed  a  pipeline  based  on Python Language (version 2.7. Available at http://www.python.org)  to  make  the  lensing  analysis. The pipeline computes the shear profile and fits a model to estimate the mass of a galaxy cluster, taking as input the observed image of the cluster. In the next subsections, we describe in detail the implemented weak lensing analysis pipeline and the results of the application on simulated data to test its performance. 
\subsection{Object detection and classification}
The first step in the lensing analysis is the detection and classification of the sources in stars and galaxies. To perform the detection and photometry of the sources we implement SExtractor \citep{Bertin96}. From SExtractor output, we use for the analysis the parameters: MAG\_BEST, as the magnitude in each filter; MU\_MAX, defined as the central surface brightness of the objects ($\mu_{MAX}$); FLUX\_MAX as the peak flux above background; FWHM as the gaussian full width at half maximum; CLASS\_STAR as the stellarity index and FLAG, which corresponds to the notes generated by SExtractor in the detection and measurement processes.

SExtractor is run twice (in a two-pass mode): A first run is made to detect bright objects in order to estimate the seeing and the saturation level of each image, and a second run to do the final detection. The first run of SExtractor is made with a detection level of 5$ \sigma $ above the background. The seeing is estimated using the average FWHM of the point-like objects selected from the FWHM/MAG\_BEST diagram, since for these objects the FWHM is independent of the magnitude. Determining the seeing is important for the star-galaxy classification, given that SExtractor uses it to compute the stellarity index. The saturation level is estimated as 0.8 times the maximum value of the FLUX\_MAX parameter. These parameters, \textit{seeing} and \textit{saturation level}, are taken into account in the SExtractor configuration file for the second run, with a lower threshold detection limit of 1.5$ \sigma $. A second run is made in dual mode, detecting objects on the $r'$ image, while astrometric and photometric pa\-ra\-me\-ters are measured on all individual images.\\
\begin{figure}
  \centering
  \includegraphics[width=.45\textwidth]{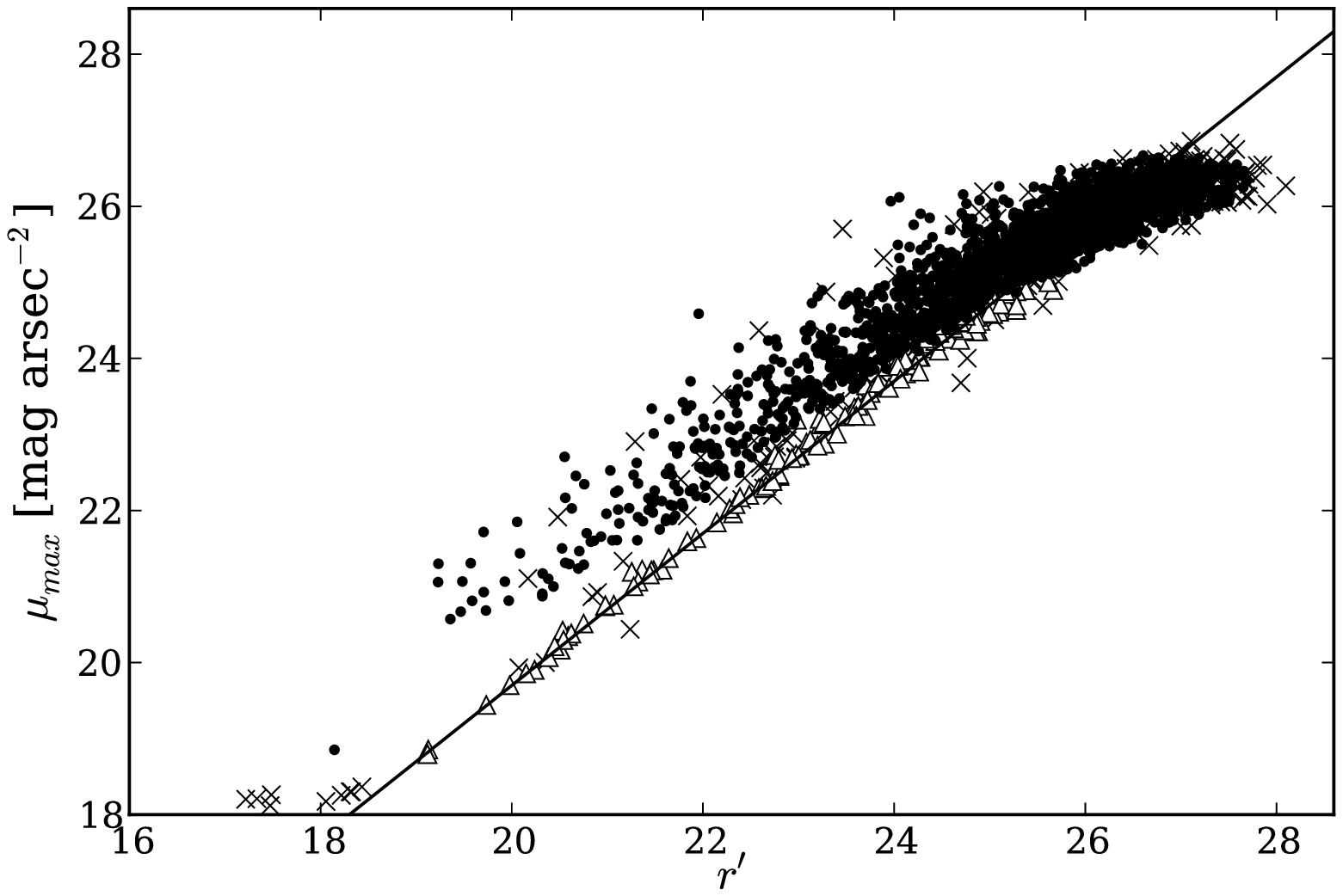}~\\
  \includegraphics[width=.45\textwidth]{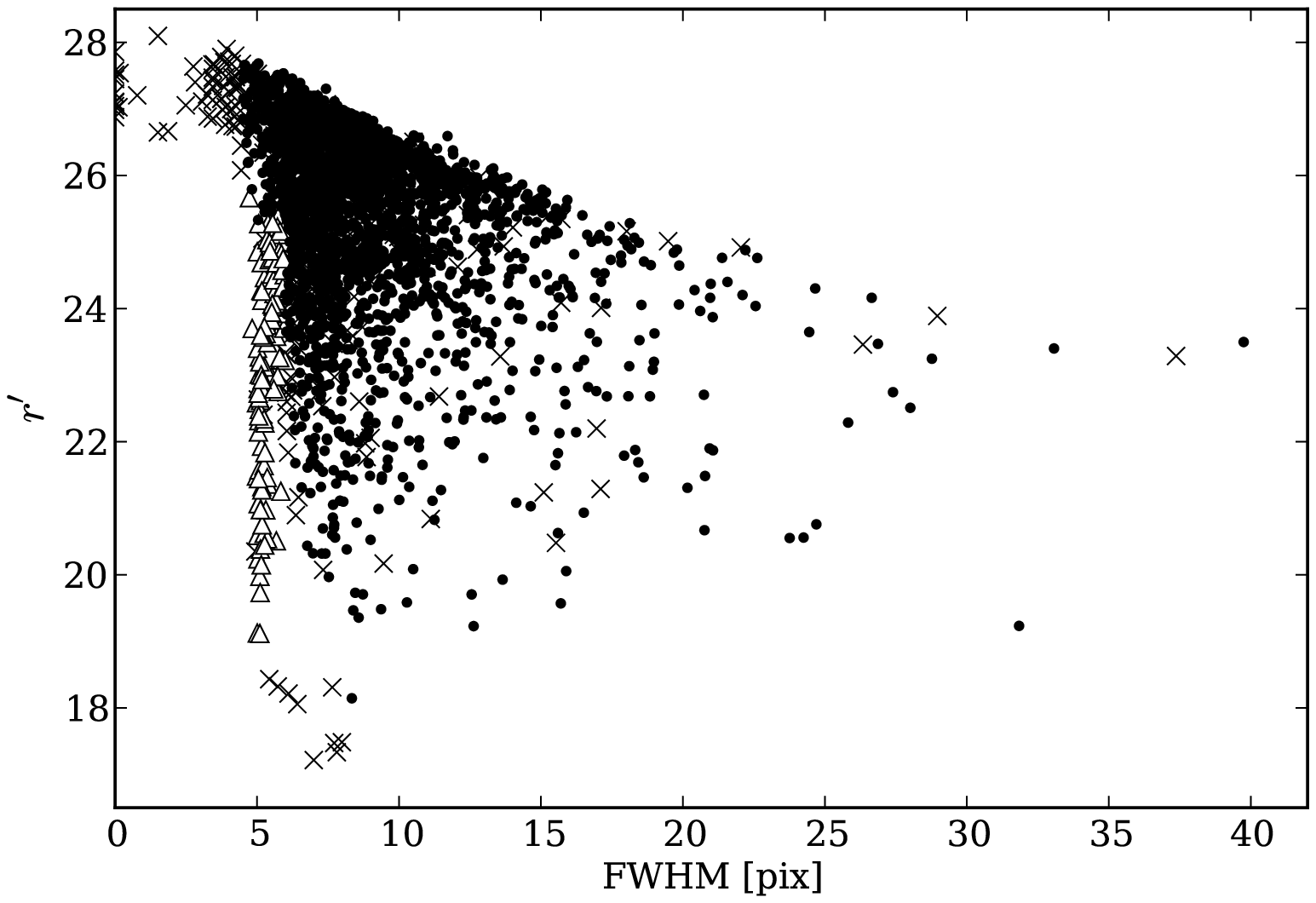}
  \caption{Classification of objects detected in the $r'$ image of the galaxy cluster [VMF98]102. Here stars  are represented by triangles, galaxies by points,  and  artifacts by cruxes. Upper pannel shows $\mu_{MAX}/r'$ plane, where stars  are  situated in  the  region  marked  by  the solid  line  $\pm$ 0.4 magnitudes, and in the lower pannel we show $r'/FWHM$ plane.}
  \label{sources}
\end{figure}
Sources are classified according to their position in  the  magnitude/central flux diagram, the FWHM respect to the seeing and the stellarity index, following \citet{Bardeau05}, in stars, galaxies and false detections. In Figure\,\ref{sources} we show, as an example, $\mu_{MAX}$ as a function of the $r'$ magnitude (u\-pper panel) and the $r'$ magnitude against the FWHM (lower panel), for all objects dected by SExtractor in the cluster [VMF098]102. Objects that are more sharply peaked than the \textit{Point Spread Function} (from now on PSF), thus with FWHM $<$ \textit{seeing} - 0.5 pixel, and with FLAG parameter $>$ 4, are considered as false detections. As the light distribution of a point source scales with magnitude, objects on the line magnitude/central flux, $\pm$ 0.4 magnitudes, FWHM\,$<$\,\textit{seeing}\,+\,1\,pixel and CLASS\_STAR $>$ 0.8 are considered as stars. The rest of the objects are considered as galaxies. \\
The first step in the pipeline ends generating two catalogues, one for the objects classified as stars and another for the galaxies.
\subsection{Shape measurements}
Measurements of galaxy shape are central in this analysis, given that galaxy ellipticities are used for the shear estimations and therefore to estimate cluster masses. It is important to take into account the roundness effects of the atmosphere as well as the distortions caused by the telescope optics, all together included in the PSF, which is convolved with the galaxy intensity distribution.

For the shape measurements we use IM2SHAPE \citep{Bridle02}. This code computes the shape parameters modeling the object as a mixture of Gaussians, convolved with a PSF which is also a sum of Gaussians. For simplicity both, the PSF and the object, are modeled  with  a  single  elliptical  Gaussian  profile. 

The PSF field across the image is estimated from the shape of the stars, since they are intrinsically point-like objects.  We only used objects with a measured ellipticity smaller than 0.2 to remove most of the remaining false detections and faint galaxies present in the catalogue. Looking at the 5 nearest stars at each position, we have also removed those that differ by more than 2$\sigma$ from the local average shape. Then,  we  linearly  interpolate  the  local  PSF  at  each object position by averaging the shapes of the five closest stars. After PSF determination, we use again IM2SHAPE to measure the galaxy shapes, and the result is a catalogue of the galaxies with its intrinsic shape parameters.

\subsection{Shear radial profiles}
\label{sec:profile}

 Gravitational lensing maps the unlensed image in the source plane, specified by coordinates $(\beta^{1},\beta^{2})$, to the lensed image $(\theta^{1},\theta^{2})$ in the image plane, using a matrix transformation:
\begin{equation*}
\left( {\begin{array}{c}
\delta\beta^1 \\
\delta\beta^2 \\
 \end{array} } \right) 
=
\left( {\begin{array}{cc}
 1-\kappa-\gamma_1 & -\gamma_2  \\
 -\gamma_2& 1-\kappa+\gamma_1  
 \end{array} } \right)
\left( {\begin{array}{c}
\delta\theta^1 \\
\delta\theta^2
 \end{array} } \right)         
;
\end{equation*}
where $\gamma_1$ and $\gamma_2$ are the components of the complex shear 
$\gamma = \gamma_1 + i \gamma_2$.
This can also be expressed as:
\begin{equation*}
\left( {\begin{array}{c}
\delta\beta^1 \\
\delta\beta^2 \\
 \end{array} } \right) 
=
(1-\kappa)
\left( {\begin{array}{cc}
 1- g_1 & - g_2  \\
 - g_2 & 1 + g_1  
 \end{array} } \right)
\left( {\begin{array}{c}
\delta\theta^1 \\
\delta\theta^2
 \end{array} } \right)         
;
\end{equation*} 
where $g_{1}$ and $g_{2}$ are the components of the reduced shear:
\begin{equation}
g=\dfrac{\gamma}{1-\kappa}
\end{equation}
which is a nonlinear function of the two lensing functions: the complex shear, $\gamma$, and the convergence, $\kappa$, related to the projected mass density.
If lensing is weak, the image of a circular source with ratio \textit{r}, appears elliptical, with axis given by
\begin{equation*}
a=\dfrac{r}{1-\kappa-|\gamma|},\,\,\,\,\,\,\,\,\,\,b=\dfrac{r}{1-\kappa+|\gamma|}
\end{equation*}
Defining the ellipticity as
\begin{equation*}
e=\dfrac{a-b}{a+b}=\dfrac{|\gamma|}{1-\kappa}\approx|\gamma|
\end{equation*}
where \textit{g} becomes the normal shear, $\gamma$, since $\kappa \ll 1$, which generally holds in the weak lensing regime for clusters, and will be assumed henceforth here.\\
If the source has an intrinsic ellipticity $\bmath{e_{s}}$, the observed ellipticity in the weak lensing limit will be:
\begin{equation*}
\bmath{e}=\bmath{e_{s}}+\gamma
\end{equation*}
Assuming that unlensed galaxies are randomly oriented on the sky plane ($\langle \bmath{e_{s}} \rangle = 0$ ) and averaging over sufficiently many sources:
\begin{equation}
\langle \bmath{e} \rangle=\langle \gamma \rangle
\end{equation}
Hence, in the weak ­lensing approximation, we get an unbiased estimator of  the  reduced  shear  by  averaging  the  shape  of  background  galaxies  in  concentric annuli  around the cluster centre. Spherical symmetry also implies that the average in annular bins of the tangential component ellipticity of the lensed  galaxies, defined as the E-­mode, traces the reduced shear. On the other hand, the average in annular bins of the component tilted at $\pi/4$  relative  to  the  tangential  component, the B-­mode, should  be exactly zero for the case of perfect symmetry \citep[e.g.][Sec.\,4]{Bartelmann01}.

Because of the random orientation of the galaxies in the source plane, the error in the observed galaxy ellip\-ti\-ci\-ties and thus, on the estimated shear, will depend on
the number of galaxies averaged together to measure the shear \citep{Schneider00}. Thus, the errors in the measured shear can be estimated as:
\begin{equation} \label{eq:err}
\sigma_{\gamma}\approx\dfrac{\sigma_{\epsilon}}{\sqrt{N}}
\end{equation}
where $\sigma_{\epsilon}$ is the dispersion of the intrinsic ellipticity distribution ($\sigma_{\epsilon} \approx 0.3$) and $N$ is the number of objects in the annular bin.

We have adopted the brightest cluster galaxy in $r'$ filter as the cluster centre, a criterium commonly used for lensing masses determinations \citep{Okabe10,Hoekstra11,Foex12}. Shear profiles were computed using nonoverlapping logarithmic annuli, in order to have similar signal-to-noise ratio (S/N) in each annuli. We have tested different annuli sizes but the final mass results have not showed a strong dependence on this parameter. We have fixed the size for the one we obtained lowest errors for the SIS and NFW profile fits. The profiles were fitted from the inner part were the signal becomes significantly positive, to reduce the impact of miscentering, up to the bin with highest number of ga\-la\-xies ($\sim$\,3\,arcmin for most of the clusters, which roughly corresponds to 0.8-1.4 Mpc). Our profiles were mainly limited by the FOV of the images. With these limits, 4-6 points were available in the shear profiles.  

\subsection{Background Galaxies selection and redshift distribution}
\label{back}
To perform the shear estimation, background galaxies were selected as those with $r'$ magnitudes between $m_{P}$ and $m_{max} + 0.5$. $m_{P}$ is defined as the faintest magnitude where the probability that the galaxy is behind the cluster is higher than 0.7 and $m_{max}$ corresponds to the peak of the magnitude distribution of galaxies in the $r'$ passband. Keeping galaxies brighter than $m_{max} + 0.5$ ensures that we are not taking into account too faint galaxies with higher uncertainties in the shape measurements. We have also restricted the objects to those with good S/N and with a good pixel sampling by using only the galaxies with $\sigma_{e} < 0.2$ ($\sigma_{e}$ is defined as the quadratic sum of the errors $\sigma_{e1}$ and $\sigma_{e2}$ given by IM2SHAPE) and with FWHM $>$ 5 pixels.  

Once we obtain a catalogue for the background ga\-la\-xies, we average the components of the ellipticities (E-mode and B-mode) in nonoverlapping annuli. The average E-mode components corresponds to the shear value which depends on the geometrical factor \mbox{$ \beta = D_{LS}/ D_{S} $}, where $D_{LS}$ is the angular diameter distance from the lens to the background source galaxy, and $D_{S}$ is the distance from the observer to the background galaxy. A galaxy at the same radial distance from the centre of the cluster but at a different background redshift is sheared differently. This variation is taken into account once we fit the profiles by $\langle\beta\rangle$.

To estimate $m_{P}$ and $\langle\beta\rangle$ we used the catalogue of photometric redshifts computed by \citet{Coupon09}, based on the public release Deep Field 1 of the Canada-France-Hawaii Telescope Legacy Survey, which is complete up to $m_{r} = 26.$. We compute the fraction of galaxies with $z >  z_{cluster}$ in magnitude bins of 0.25 magnitudes for the $r'$ filter, and then we chose $m_{P}$ as the lowest magnitude for which the fraction of galaxies was greater than 0.7. Then we applied the photometric selection criteria to the catalogue ($m_{P} < m_{r} < m_{max} + 0.5$) and we computed $\beta$ for the whole distribution of galaxies. To take into account the contamination by foreground galaxies given our selection criteria, we set $\beta(z_{phot} < z_{cluster}) = 0$  which outbalances the dilution of the shear signal by these unlensed galaxies. Deep Field\,1 covers a sky region of 1\,degree$^2$, thus to estimate the cosmic variance, we divide the field in 25 non-overlapping areas of $\sim$\,140\,arcmin$^2$ and we compute $m_{P}$ and $\langle \beta \rangle$ at $z_{cluster}=0.5$ for each area. The uncertainties due to the cosmic variance were estimated as the dispersion of the values obtained for each area, obtaining $\sim 0.3$ for $m_{P}$ and $\sim 0.01$ for $\langle \beta \rangle$. Given that the errors in $\langle \beta \rangle$ are lower than the $3\%$, which represents an error of the $\sim\,5\%$ in the mass, we did not consider these uncertanties in the estimation of the masses errors since the uncertainty due to the intrinsic shape of field galaxies is much bigger.

\begin{figure}
\centering
\includegraphics[scale=0.45]{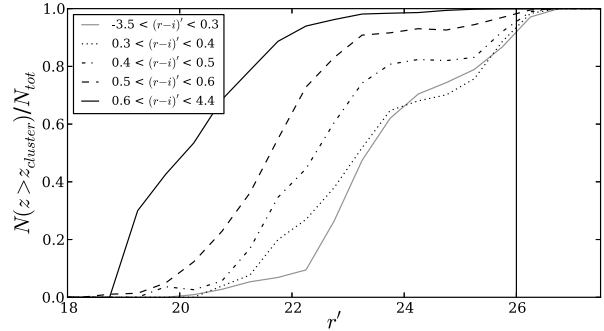}
\caption{Fraction of galaxies with $z >  0.5$ ($N(z>0.5)/N_{tot}$), for different magnitudes in filter $r'$ and colours \textit{r'-i'}, computed using photometric redshifts given by \citet{Coupon09}, used to compute the weight for the shear estimation. The vertical line idicates the $m_{max}$ position (see text for its definition).}
\label{weigh}
\end{figure}

In order to take into account the contamination of foreground galaxies in the catalogue, we weighted the estimated shear, $\langle \gamma \rangle$, with the probability that the galaxy was behind the cluster. We compute this probability using \citeauthor{Coupon09} catalogue, from the fraction of galaxies with $z >  z_{cluster}$ for each bin in magnitude, $r'$, and colour (\textit{g' - r'} and \textit{r' - i'}), see Figure\,\ref{weigh}. Hence, given the magnitude and the colour of each galaxy, we assigned to it a weigh, \textit{w}, as the fraction of galaxies with $z >  z_{cluster}$ in that bin. For [VMF98]102 we have only one image in the filter $r'$, therefore for weighing the shear profile we take into account the probability that each galaxy was behind the cluster given the magnitude of that galaxy (we did not take into account the colours for computing this probability, as in the other clusters).

\subsection{Fitting the profiles} 

\begin{figure*}
  \includegraphics[scale=0.4]{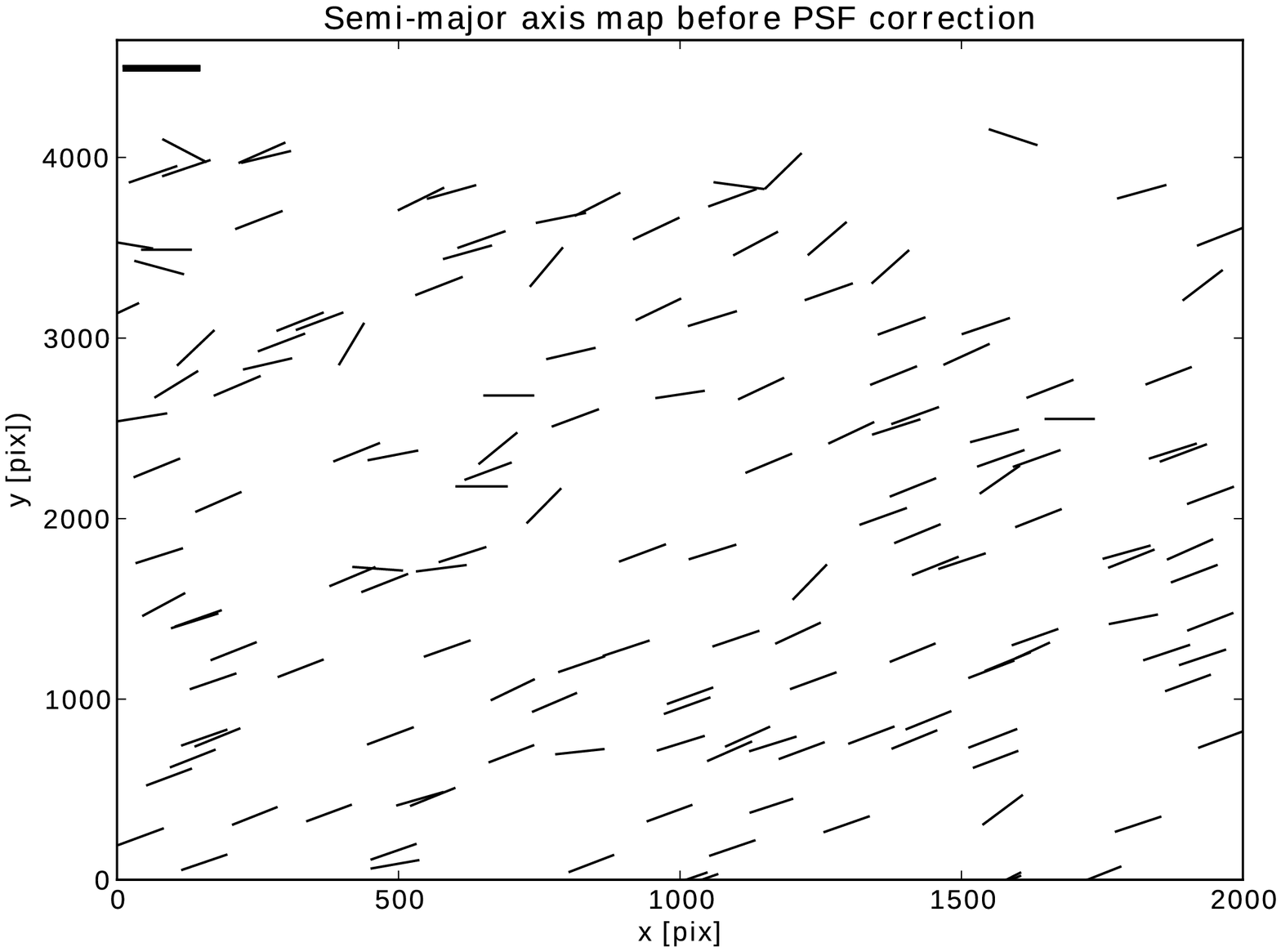}~\hfill
 \includegraphics[scale=0.4]{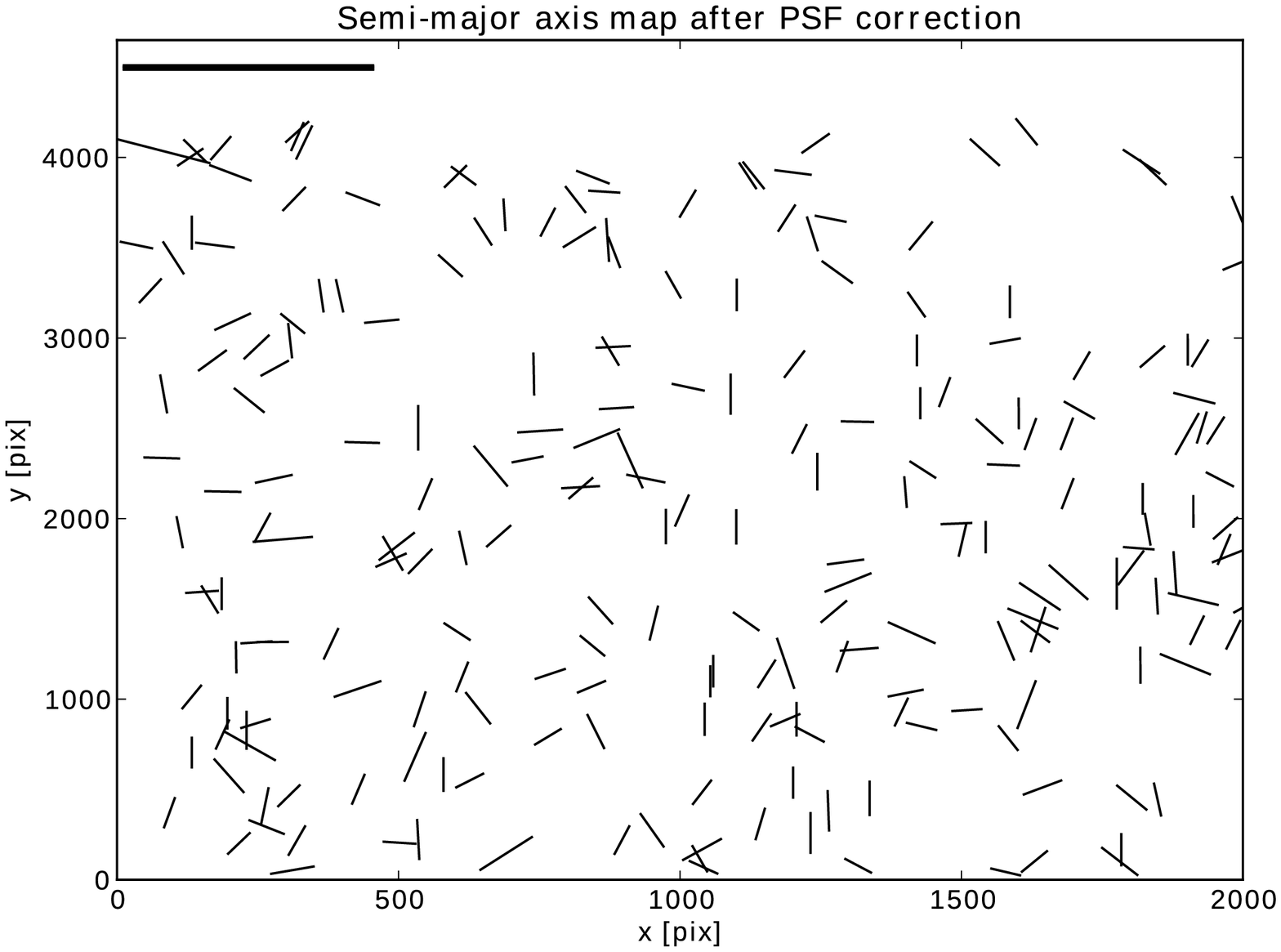}
  \caption{PSF treatment applied to stars of one of the images of the DES 
simulation: Mayor semiaxis ($a$ cos $\theta$, $a$ sin $\theta$) before (\textit{left}) and after (\textit{right}) the PSF deconvolution in the CCD. Notice that the semiaxis are more randomly distributed and the scale (given by the first thicker segment in the upper-left corner and which corresponds to 3 pix) is much more smaller after the taking into account the PSF. }
 \label{PSF}
\end{figure*}

\begin{figure*}

\centering
\includegraphics[scale=0.44]{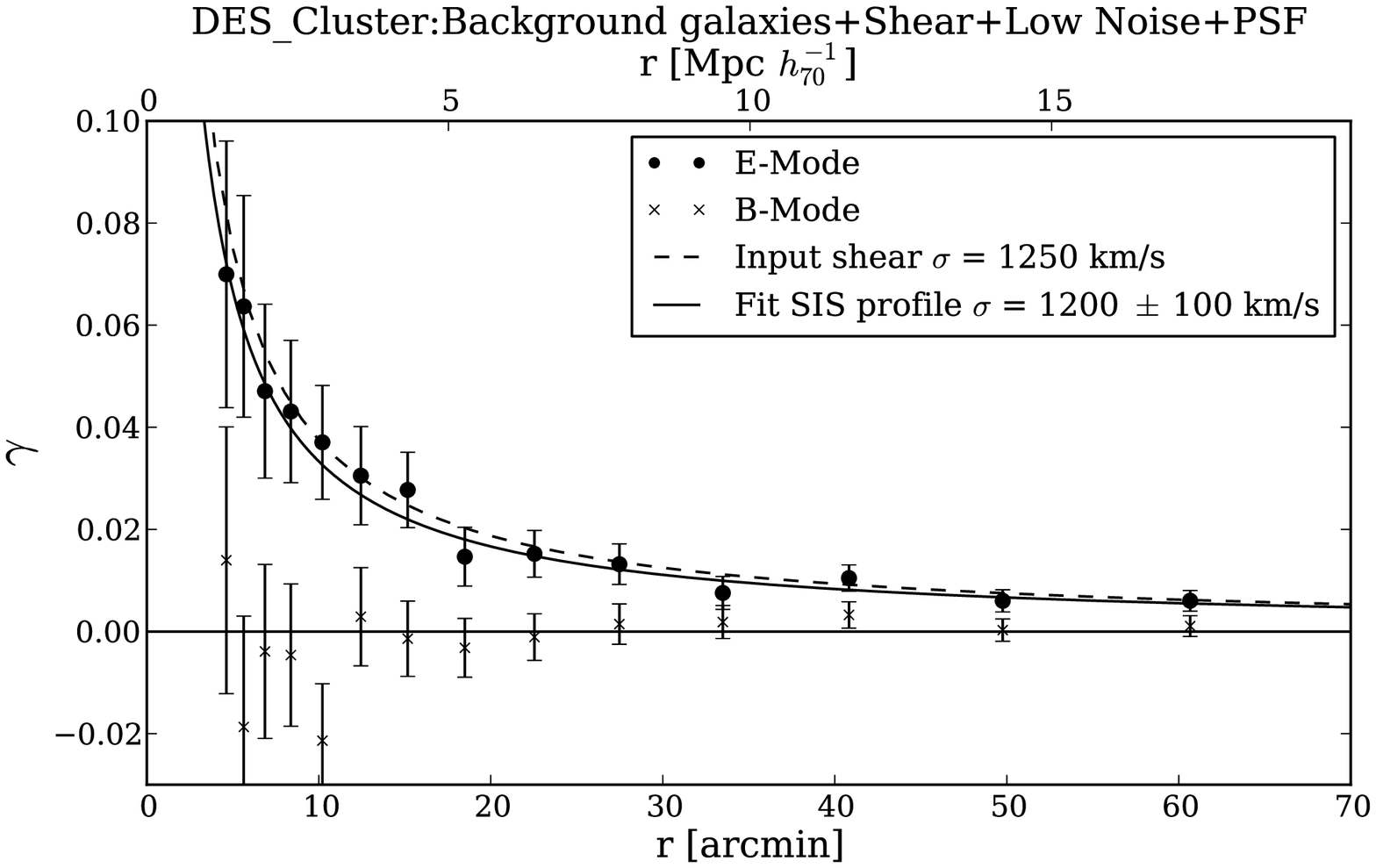}\hfill
\includegraphics[scale=0.44]{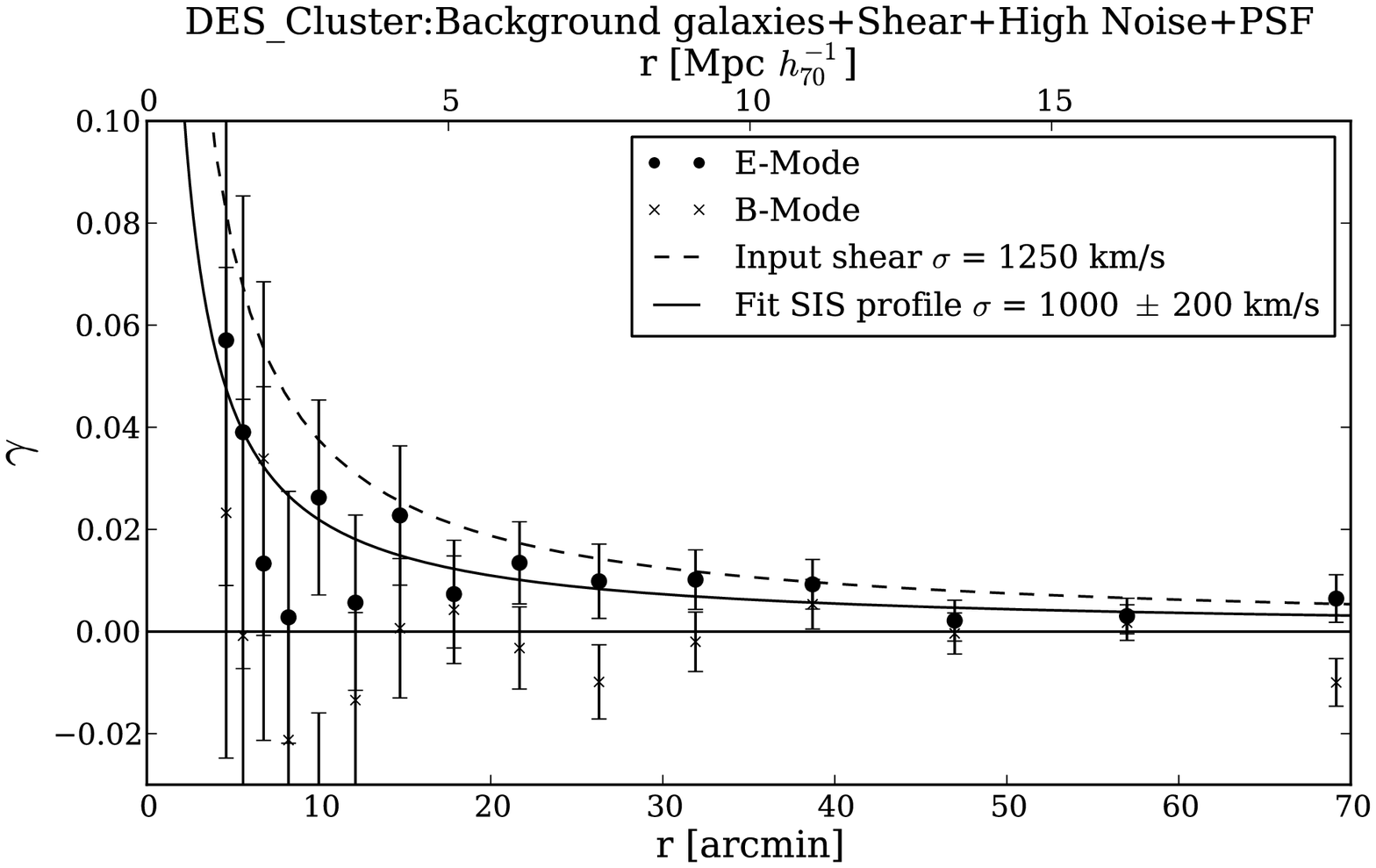} 

\caption{Shear profiles obtained for the Low (\textit{left}) and High (\textit{right}) Noise PSF Applied File, from the DES cluster simulation. The dashed curve shows the SIS profile for the input value of $\sigma_{V}$ and the solid one the fitted profile. E and B modes are represented by full circles and crosses, respectively.}
\label{DES}

\end{figure*}  

 We finally estimate the M$_{200}$ mass, defined as \mbox{M$_{200}\equiv\,$M$\,(\,<\,$R$_{200})\,=200\rho_{crit}(z)\dfrac{4}{3}\pi\,r_{200}^{3}$}, where $R_{200}$ is the radius that encloses a mean density equal to 200 times the critical density ($\rho_{crit} \equiv 3 H^{2}(z)/8 \pi G$; $H(z)$ is the redshift dependent Hubble parameter and $G$ is the gravitational constant). In order to do that we fit the shear data with the singular isothermal sphere (SIS) and the NFW profile \citep{Navarro97} using $\chi^{2}$ minimization. These density profiles are the standard parametric models used in lensing analysis to characterize the lenses. Following, we explain briefly the lensing formulae for these two profiles:

\subsubsection{SIS profile}

The SIS mass model is the simplest one for describing a relaxed massive sphere with a constant and isotropic ve\-lo\-ci\-ty dispersion. This is mainly described by the density distribution:
\begin{equation*}
\rho(r) =  \dfrac{\sigma_{V}^{2}}{2 \pi G r^{2}}
\end{equation*}
This model corresponds to a distribution of self-gravitating particles where the velocity distribution at all radii is a Maxwellian with one dimensional velocity dispersion, $\sigma_{V}$. From this equation, we can get the critical Einstein radius for the source sample as:
\begin{equation}\label{eq:SIS}
\theta_{E} = \dfrac{4 \pi \sigma_{V}^{2}}{c_{vel}^{2}} \frac{1}{\langle \beta \rangle}
\end{equation}
where $c_{vel}$ is the speed of light, in terms of wich one obtains:
\begin{equation}
\kappa_{\theta} = \gamma_{\theta} = \dfrac{\theta_{E}}{2 \theta}
\end{equation}
where $\theta$ is the distance to the cluster centre. Hence, fitting the shear for different radius, we can estimate the Einstein radius, and from that, we can obtain an estimation of the mass M$_{200}$ as \citep{Leonard10}:
\begin{equation}\label{eq:MSIS}
M_{200} =  \dfrac{2 \sigma_{V}^{3} }{\sqrt{50} G H(z)} 
\end{equation}

\subsubsection{NFW profile}
The NFW profile is derived from fitting the density profile
of numerical simulations of cold dark matter halos \citep{Navarro97}. This profile depends on two parameters, the virial radius, R$_{200}$, and a dimensionless concentration parameter, \textit{c}:
\begin{equation*}
\rho(r) =  \dfrac{\rho_{c} \delta_{c}}{(r/r_{s})(1+r/r_{s})^{2}} 
\end{equation*}
where $r_{s}$ is the scale radius, $r_{s} = $R$_{200}/c$ and $\delta_{c}$ is the cha\-rac\-te\-ris\-tic overdensity of the halo,
\begin{equation*}
\delta_{c} = \frac{200}{3} \dfrac{c^{3}}{\ln(1+c)-c/(1+c)}  
\end{equation*}
We used the lensing formulae for the spherical NFW density profile from \citet{Wright00}. If we fit the shear for different radius we can have an estimation of the pa\-ra\-me\-ters $c$ and R$_{200}$. Once we obtain R$_{200}$ we can compute the M$_{200}$ mass. Nevertheless, there is a well-known degeneracy between the parameters R$_{200}$ and $c$ when fitting the shear profile in the weak lensing regime. This is due to the lack of information on the mass distribution near the cluster centre and only a combination of strong and weak lensing can raise it and provide useful constraints on the concentration parameter. Since we do not have strong lensing modeling for the clusters in the sample, we decided to fix the concentration parameter, $c_{200} = 4$, according to the predicted concentrations given by \citet{Duffy11} for a relaxed cluster with M$ = 1 \times 10^{14} M_{\odot} h_{70}^{-1}$ placed at $z \sim 0.4$. Thus, we fit the mass profile with only one free parameter, R$_{200}$.

\subsection{Testing the pipeline with simulated data}
\label{sec:sim}

To check the performance of our weak lensing analysis pipeline, we tested it on the DES cluster simulation images publically available \citep{Gill09}. This simulation consists of a sets of images, with different grades of difficulty, of sheared galaxies due to the presence of a SIS profile with a velocity dispersion of 1250\,km\,s$^{-1}$. This is a suitable test for our pipeline given that the idea is to apply it to real clusters of galaxies. We applied our pipeline to three of the available image files, High Noise File, High Noise PSF Applied File and Low Noise PSF Applied File.

For the PSF Applied files, we checked that our IM2SHAPE implementation can recover point-like objects by applying the PSF correction to each star. Figure\,\ref{PSF} shows the results of the shape parameters measurements for these stars, with and wihtout taking into account the PSF in the shape measurement: the size distribution is dominated by point sources, and the orientation is more uniformly distributed after the PSF correction.

The images contain only the sheared galaxies, hence all the galaxies detected were considered as background galaxies at $z=0.8$, which is the average redshift of the galaxies. We cut the catalogue discarding the galaxies with \mbox{FWHM $<$ 5} and with $\sigma_{e} > 0.2$. Shear profiles are shown in Figure\,\ref{DES}. For the most complex image that we treated (high noise image of sheared galaxies convolved with a PSF), we obtained a deviation parameter of 1.3, defined as the number of $\sigma$ that the result is away from the input value of $\sigma_{V} = 1250$ km/sec, i.e. $\sigma = \dfrac{result - input}{error}$ , where the e\-rror was estimated according to the root mean square e\-rror of the Einstein radius. Given these results, we conclude that our weak lensing pipeline is able to reproduce the input shear signal, thus it could be applied to real observations to extract the lensing signal and to estimate the masses of cluster of galaxies.

\begin{table*}

\caption{Main results of the weak lensing analysis}\label{tab:esp}
\label{table:2}

\begin{tabular}{@{}crrrrccccrrcr@{}}
\hline
\rule{0pt}{1.05em}%
   [VMF\,98]  &  $\alpha$ & $\delta$ & $\rho_{back}$ & $m_{P}$ & $m_{max}$ & $\langle\beta\rangle$&$\sigma_{V}^{spec}$ & \multicolumn{2}{c}{SIS} & \multicolumn{2}{c}{NFW}   \\
   Id. & (J2000) & (J2000) &    & & & &  & $\sigma_{V}$ & M$_{200}$ & R$_{200}$ & M$_{200}$   \\

 \hline
\rule{0pt}{1.05em}%
001  &  00	30	34.0	&   +26  18  10	    &  56 &  23.0 & 26.1 & 0.41  & -    & 780   $\pm$ 100  &    3.4 $\pm$    1.3  &   1.3  $^{+0.2}_{-0.2}$ &   4.0$^{+2.2}_{-2.0}$ \\ 
022  &  02	06	21.2	&	+15	 11	 01	   	&  18 &  20.7 & 25.1 & 0.61  &  508 & 570   $\pm$ 100  &    1.5 $\pm$    0.8  &   1.1  $^{+0.2}_{-0.2}$ &   2.1$^{+1.2}_{-1.1}$ \\ 
093  &  10	53	18.9	&	+57	 20	 45	   	&   8 &  22.3 & 24.0 & 0.48  &   -  & 750   $\pm$ 140  &    3.4 $\pm$    1.9  &   1.4  $^{+0.4}_{-0.4}$ &   4.0$^{+3.6}_{-3.1}$ \\ 
097  &  11	17	26.1	&	+07	 43	 35	   	&  40 &  23.0 & 26.0 & 0.43  &  775 & 720   $\pm$ 100  &    2.7 $\pm$    1.1  &   1.1  $^{+0.3}_{-0.2}$ &   2.8$^{+1.9}_{-1.7}$ \\ 
102  &  11  24  05.8	&   -17  00  50    	&  40 &  22.7 & 25.9 & 0.49  &  675 & 650   $\pm$ 120  &    2.1 $\pm$    1.2  &   1.2  $^{+0.3}_{-0.2}$ &   2.7$^{+1.9}_{-1.7}$ \\ 
119  &  12	21	29.3	&	+49	 18	 40	   	&  13 &  24.5 & 25.4 & 0.29  &  -   & 1000   $\pm$ 160  &    6.3 $\pm$    3.1  &   1.4  $^{+0.2}_{-0.2}$ &   7.3$^{+3.8}_{-3.4}$ \\ 
124  &  12	52	04.1	&	-29	 20	 29  	&  33 &  19.5 & 25.7 & 0.71  &  700 & 430   $\pm$  60  &    0.7 $\pm$    0.3  &   0.8  $^{+0.3}_{-0.2}$ &   0.8$^{+0.8}_{-0.7}$ \\ 
148  &       -          &        -          &  26 &  24.5 & 25.9 & 0.29 &  -    & -              & -              & -                 & \\

\hline         
\end{tabular}
\medskip
\begin{flushleft}
\textbf{Notes.} Columns: (1) shows the cluster identification; (2) and (3), the coordinates of the centre adopted for the lensing analysis; (4), the density of background galaxies (galaxies\,arcmin$^{-2}$); (5) and (6), the brightest and faintest magnitude limits considered for the galaxy
background selection (see Section\,\ref{back}); (7), the geometrical factor; (8), the line-of-sight spectroscopic velocity dispersion from Paper\,I; (9) and (10) the results from the SIS profile fit, the velocity dispersion and M$_{200}$ (see Equations\,\ref{eq:SIS} and \ref{eq:MSIS}); (11) and (12), the results from the NFW profile fit, R$_{200}$ and M$_{200}$. The velocity dispersion, M$_{200}$ and R$_{200}$ are in units of km\,s$^{-1}$,$10^{14} M_{\odot} h_{70}^{-1}$ and Mpc\,$h_{70}^{-1}$, respectively.
\end{flushleft}
\end{table*}

%
\section{RESULTS}
\label{sec:results}

\begin{figure*}
\centering
\includegraphics[scale=0.6]{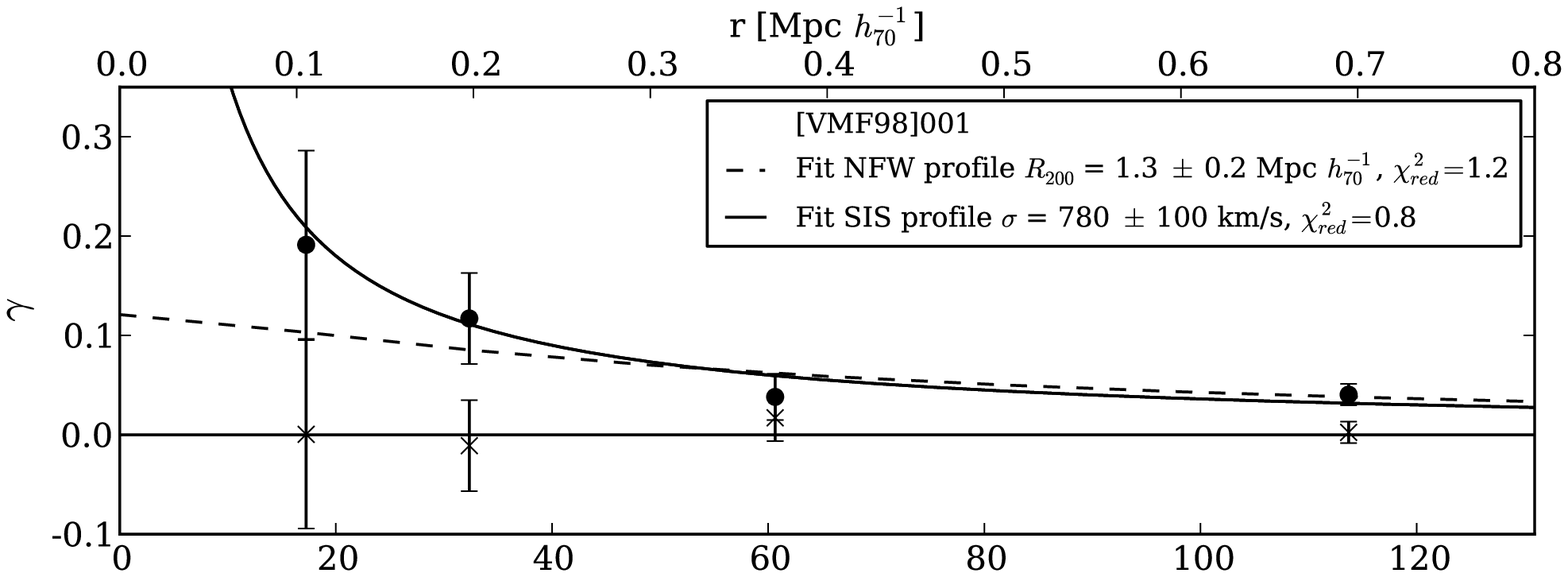} 
\includegraphics[scale=0.6]{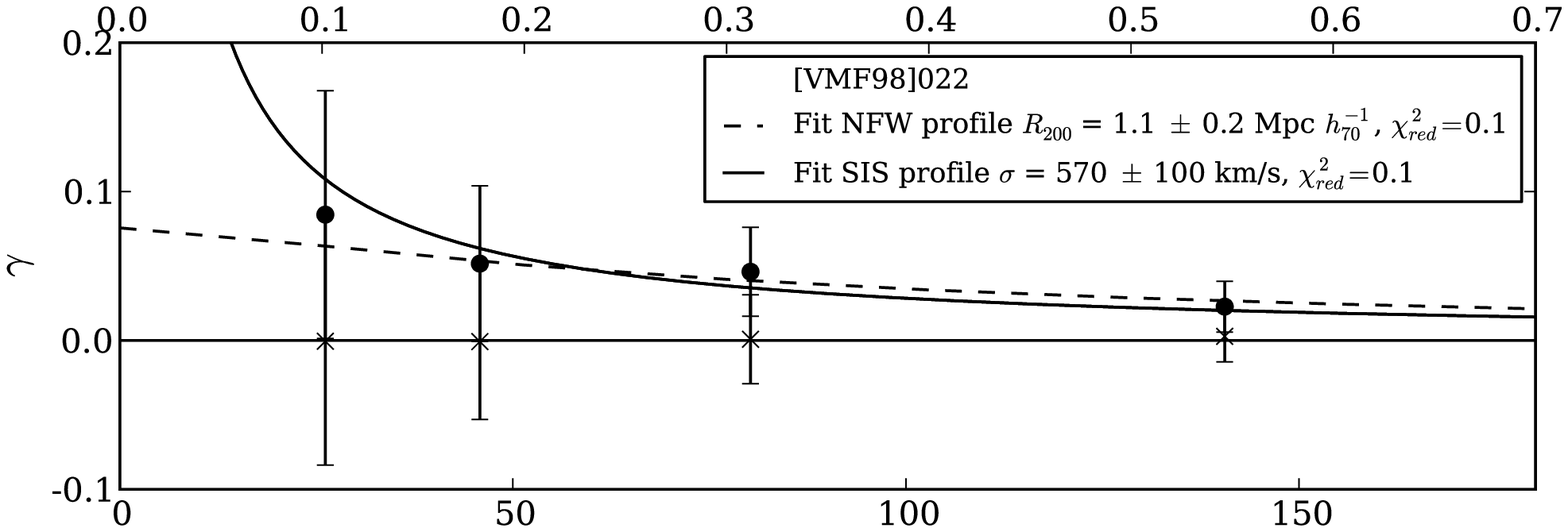}
\includegraphics[scale=0.6]{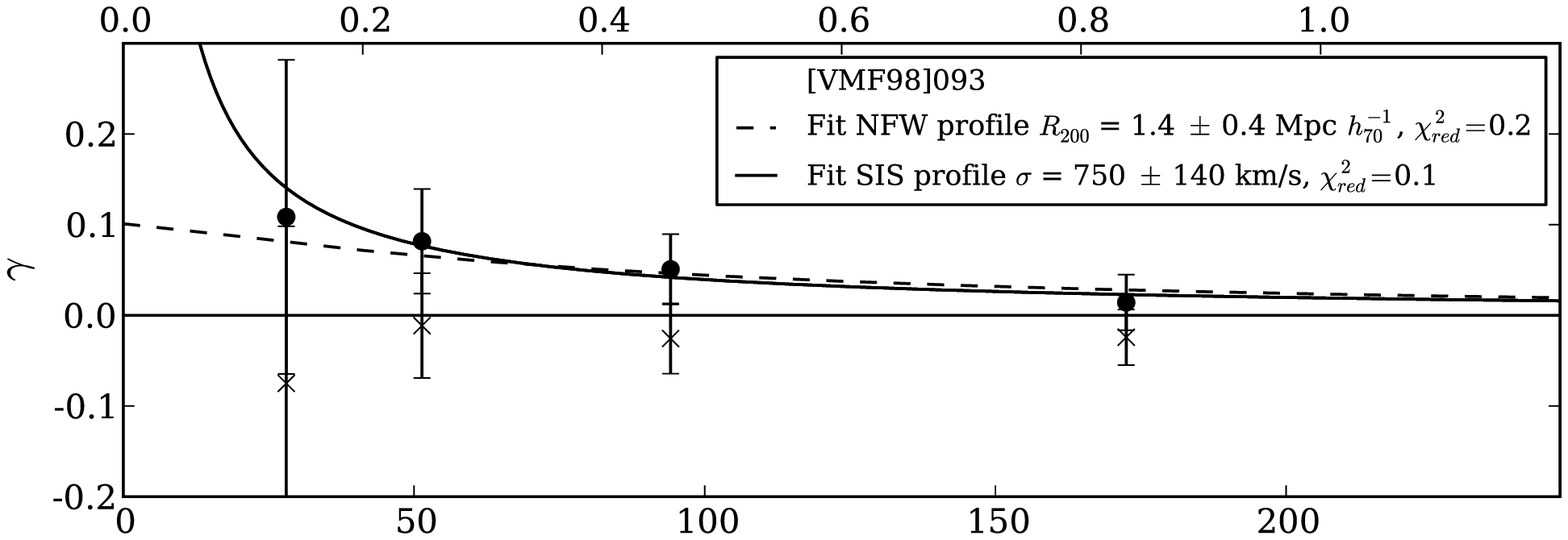}
\includegraphics[scale=0.6]{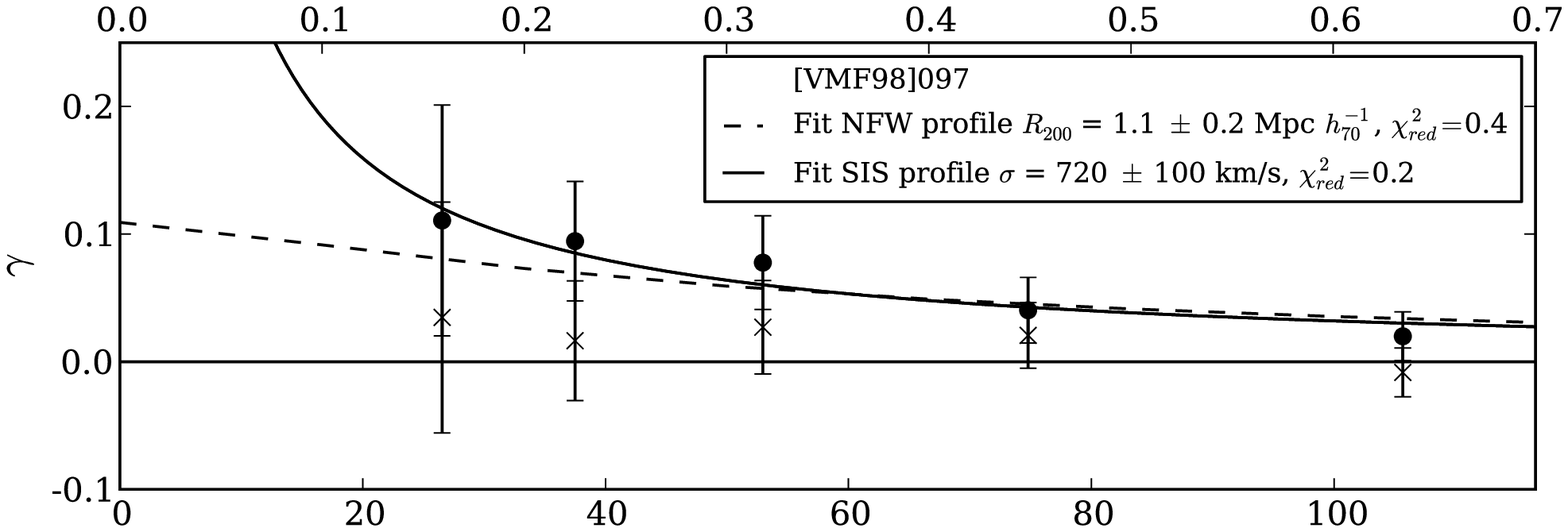}
\includegraphics[scale=0.6]{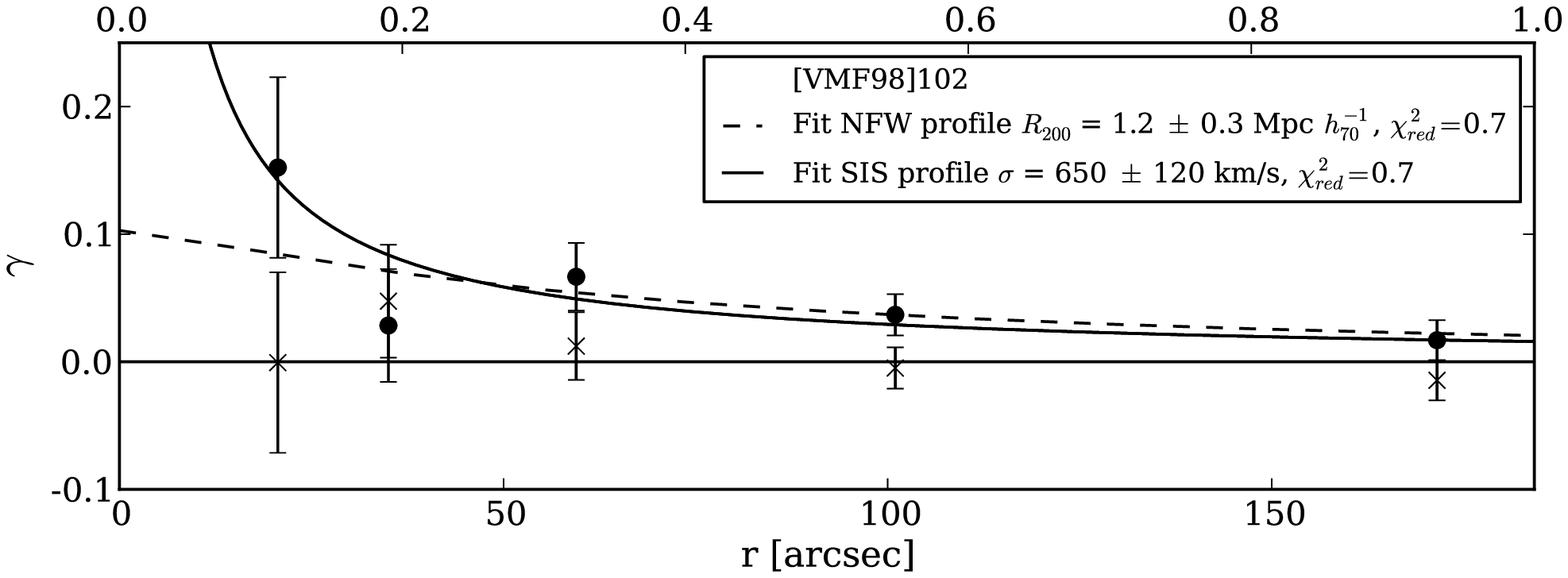} 
\caption{Shear radial profiles as a function of cluster-centric projected distance (in arcsec and Mpc) obtained for the $r'$ images of sample of clusters. The solid and the dashed lines represent the best fit of SIS and NFW profiles, respectively, with the fitted parameters given in the box. The points and crossings show the E­ and B­ modes profiles averaged in annular bins, respectively. Error bars are computed according to Equation\,\ref{eq:err}.}
\label{shear-profile1}
\end{figure*}

\begin{figure*}
\centering
\includegraphics[scale=0.6]{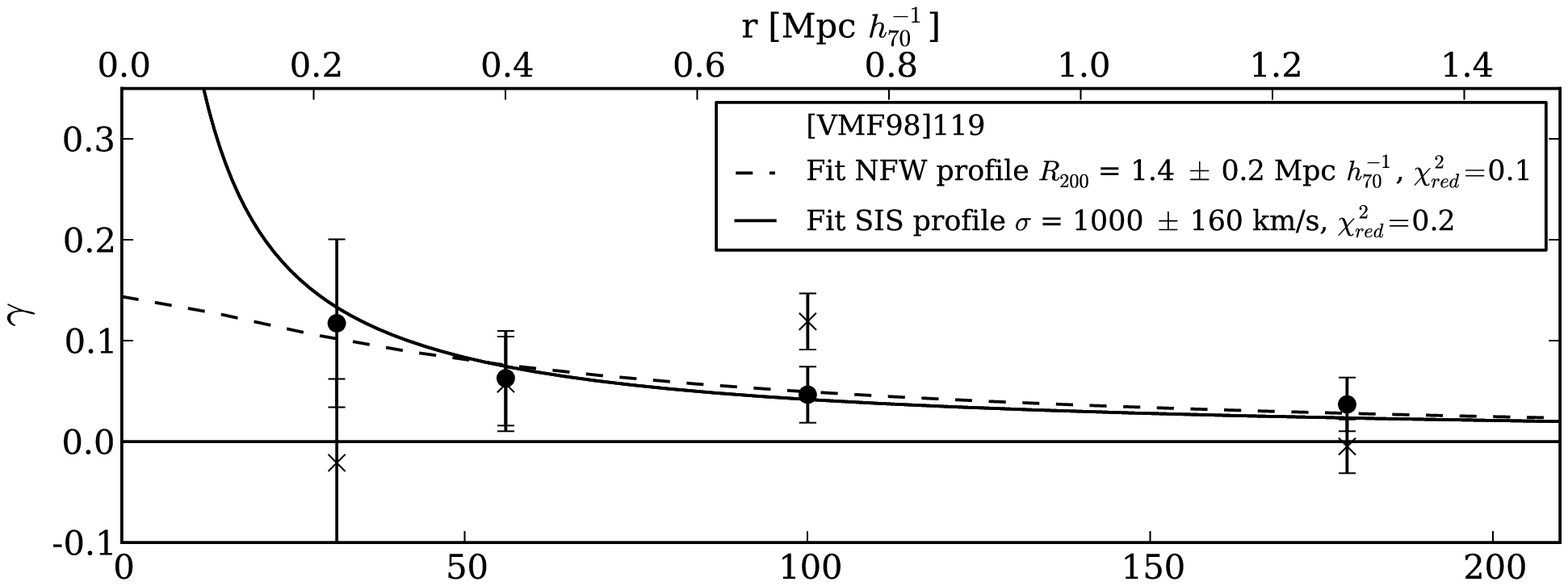} 
\includegraphics[scale=0.6]{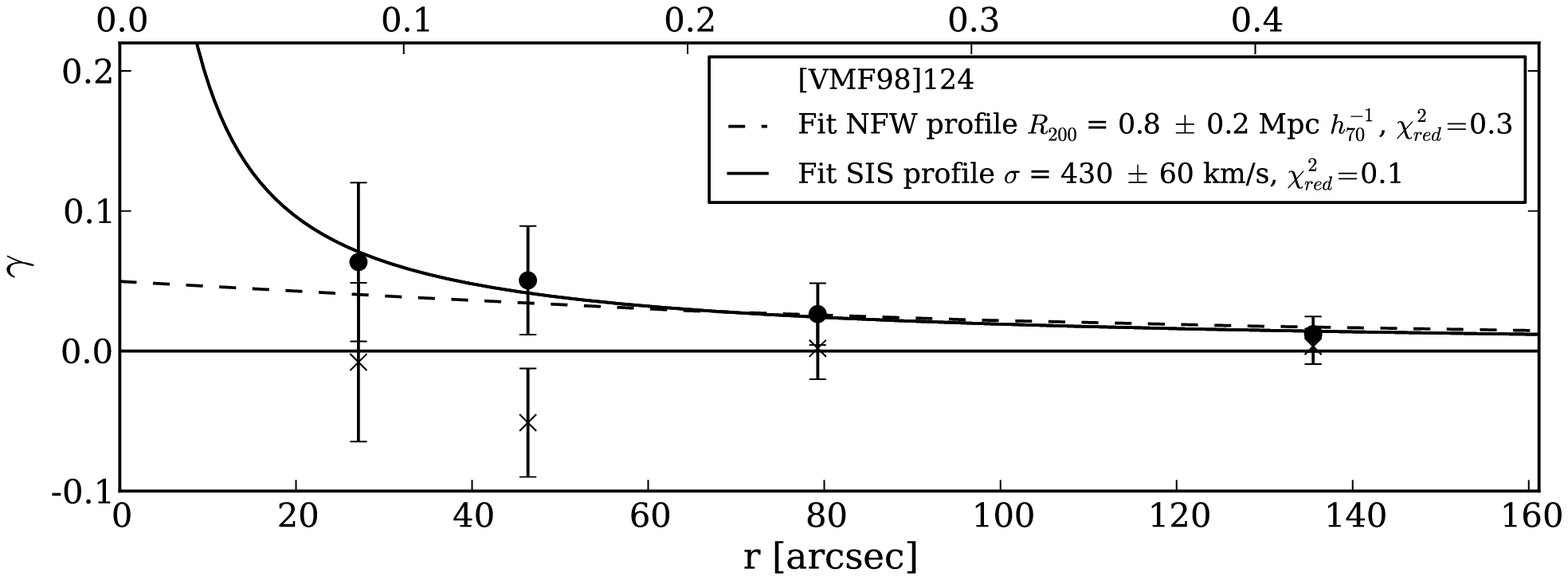} 
\caption{Shear radial profiles as a function of cluster-centric projected distance (in arcsec and Mpc) obtained for the $r'$ images of sample of clusters. The solid and the dashed lines represent the best fit of SIS and NFW profiles, respectively, with the fitted parameters given in the box. The points and crossings show the E­ and B­ modes profiles averaged in annular bins, respectively. Error bars are computed according to Equation\,\ref{eq:err}.}
\label{shear-profile2}
\end{figure*}

From our weak lensing analysis, we estimated the mass of seven clusters in the sample. Due to its low signal to noise, for cluster [VMF98]148 it was not possible to derive a reliable mass estimate from our lensing measurements. The results of the analysis are shown in Table\,\ref{table:2}. Errors in $\sigma_{V}$, $R_{200}$ and the masses were computed according to the $\chi^{2}$ dispersion. Errors in M$_{200}^{NFW}$ are higher than M$_{200}^{SIS}$, given the big uncertainties in the R$_{200}$ parameter . Nevertheless, both estimations are consistent being the NFW masses sys\-te\-ma\-ti\-ca\-lly larger by a $\sim 20 \%$ ($\langle$ M$_{200}^{NFW} / $ M$_{200}^{SIS} \rangle$ = 1.21 $\pm$ 0.13, where the uncertainty corresponds to the scatter around the mean), in excellent agreement with the result presented by \citet{Okabe11} for the virial masses. Shear profiles obtained for the galaxy clusters are shown in Figures\,\ref{shear-profile1} and \ref{shear-profile2} with the reduced $\chi^{2}$ for each fit. We include both fits, SIS (solid line) and NFW (dashed line) models. Points and crosses represent the E and B ­modes averaged in annular bins, respectively. All profiles are well fitted by both mo\-dels. In the next subsections we discuss our results and we study the relation between the mass derived and the cluster X-ray luminosities. 

\subsection{Properties of individual clusters}

\subsubsection{[VMF]001}

For the galaxy cluster [VMF98]001 we obtained a shear signal consistent with a velocity dispersion of $\sim$\,800\,km\,s$^{-1}$. There is a big offset between the position BCG, adopted as the centre
for the lensing analysis, and the X-ray luminosity peak from ROSAT ($\sim$\,110\,kpc), not observed in the X-ray countours obtained with XMM-Newton (see Figure 6, from Paper I). Thus, given the lower resolution of ROSAT observations, the X-ray peak might be poorly determinated leading to unrealistic offsets. Further evidence of this fact is the absence of the shear profile signal centred at the X-ray position. We argue that the centre of the gravitational potential should be close to the BCG position.

\subsubsection{[VMF]022}

The galaxy cluster [VMF98]022 shows an elongated distribution of galaxies in the NE-SW direction. The cluster is dominated by a bright elliptical galaxy, which presents a shift of $\sim$12$^{\prime\prime}$ in the south-west direction with respect to the X-ray peak emission (for further details about the cluster morphology, see Section\,3.4 in Paper II). We compute the shear profile taking this bright elliptical as the centre of the cluster. This system presents a shear profile signal consistent with a velocity dispersion of 540\,km\,s$^{-1}$, in good agreement with the velocity dispersion fitted from the redshift distribution (see Sec.\,4 in Paper\,I).

\subsubsection{[VMF]093}

For the cluster [VMF98]093, in spite of the low density of background galaxies, we obtain a significant signal consistent with a velocity dispersion of 750\,km\,s$^{-1}$. As evidence of the relaxed state of this cluster, we observe a dominant population of red galaxies as well as concentric X-ray countours centred in the BCG (Figure 6 in Paper I).

\subsubsection{[VMF]097}

The galaxy cluster [VMF98]097 was previously a\-na\-ly\-sed by \citet{Carrasco07}, using the same set of ima\-ges. They obtained a large discrepancy between mass estimates, where the X-ray mass exceeds by more than a factor three the weak lens derived estimate. Moreover, they found a large degree of substructure, as also seen in the redshift distribution presented in Paper I (Figure 11). However, substructure cannot explain the defect in the weak lensing mass, given that substructure in the surroundings would tend to dilute the tangential shear leading to mass under-estimation \citep{Meneghetti10,Giocoli12,Giocoli14}. We improve the profile \citep[see Figure\,9 from][p. 11]{Carrasco07}, adding a new constraint for the E-mode and obtain a profile consistent with zero for the B-mode. Nevertheless, our weak lensing mass estimate is consistent with that obtained by \citet{Carrasco07}, corresponding to a velocity dispersion of $\sim$\,700\,km\,s$^{-1}$ and also, with the velocity dispersion from the redshift distribution of 775\,km\,s$^{-1}$ (see Sec.\,4 in Paper\,I).

\begin{figure*}
\centering
\includegraphics[scale=0.78]{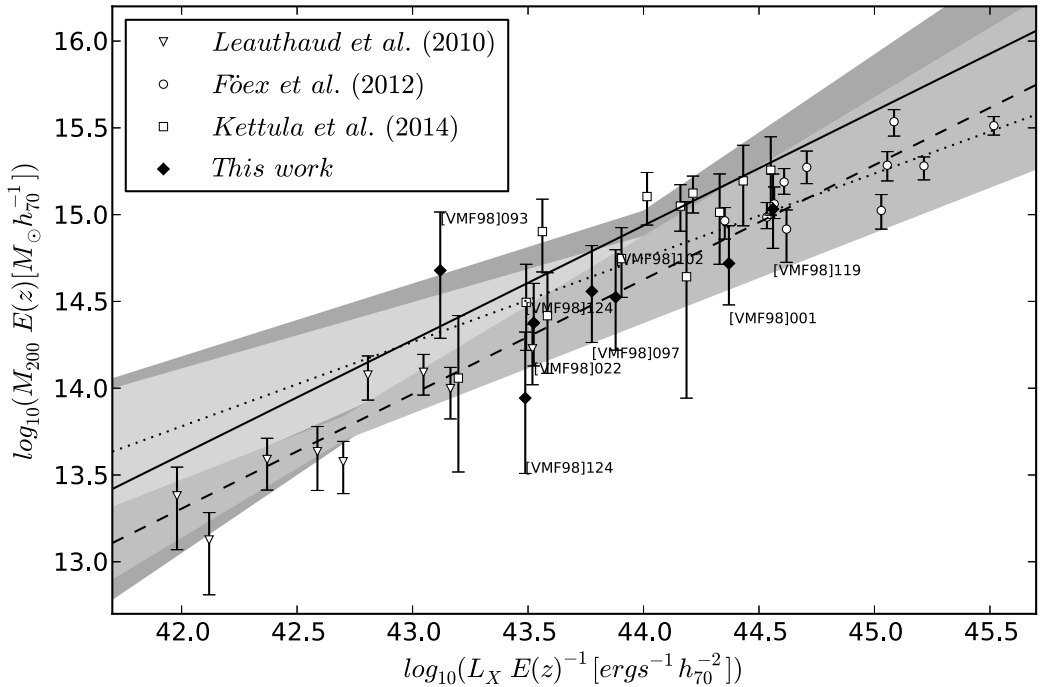}
\caption{Weak lensing masses versus X-ray luminosities for the sample of clusters (diamonds), combined with the stacked measurement by \citet{Leauthaud10} (open triangles), EXCPRES clusters by \citet{Foex12} (open circles) and low-mass from the CFHTLS (open squares) by \citet{Kettula14}. Dashed, pointed, and solid lines represent the fit obained by \citet{Leauthaud10}, \citet{Foex12} and \citet{Kettula14}, respectively.}
\label{lumass}
\end{figure*}

\subsubsection{[VMF]102}

The results from cluster [VMF98]102 give a velocity dispersion of 640\,km\,s$^{-1}$  , in good agreement with the spectroscopic value obtained in Paper I. In this case, the profile was built adopting a centre between the X-ray peak and the second brightest galaxy member. This was selected af\-ter trying to fit the profile taken the centre as the second brightest galaxy and then, as the X-ray peak, without ge\-tting enough signal-noise ratio to fit the profile. The second brightest galaxy in this case is close to the other bright cluster members and, unlike the brightest galaxy, it is an elliptical galaxy, so it is a more adequate guess for the cluster centre in this case. This cluster presents irregular X-ray contours (Figure\,6 in Paper\,I) and, based in spectroscopic information, we found a non related group of galaxies in the line of sight (Figure\,11, in Paper\,I). Also, there is a big offset between the X-ray peak and the centre adopted for the lensing analysis ($\sim$\,220\,kpc), however we could not confirm this offset with higher resolution observations.

\subsubsection{[VMF]119}

The cluster [VMF98]119 is one of the highest redshift clusters ($ z \sim 0.7 $) in our sample. Even with a very low density of background galaxies, it shows a significant shear signal according with a velocity dispersion of 1000\,km\,s$^{-1}$. The centre was placed at the brightest galaxy member, $\sim$\,1' from the ROSAT X-ray peak. Using X-ray observations from CHANDRA Data Archive, we built the X-ray contours and the peak is displaced $\sim$0.9' from the ROSAT centre, but still $\sim$0.4' ($\sim$170\,kpc) displaced from the BCG. Also, the B-modes do not follow a null flat profile, which could be suggesting a large deviation from the spherical symmetry. This can also be seen in the distribution of member galaxies (Figure\,12 in Paper\,II).

\subsubsection{[VMF]124}

Finally, for the cluster [VMF98]124, the centre from the X-ray data using XMM-Newton contours (Figure\,6 in Paper\,I) agrees with the BCG position. Besides, there is no evidence of another group in redshift space (see Figure\,11 from Paper\,I) and we observed a dominant red galaxy population (see Paper\,II), which indicates the relaxed state of this system. This cluster presents a low shear signal consistent
with 430\,km\,s$^{-1}$. There is a large difference between the velocity dispersion obtained by the lensing analysis and that derived from the redshift distribution (700\,km\,s$^{-1}$, Sec.\,4 in Paper I).  We notice, however, the high uncertainty in this value given the small number of available redshifts.  

\subsection{$M - L_{X}$ relation}
\label{lmrelation}

We have also investigated the relation between the estimated mass and the X-ray luminosity, which is
a diagnostic of the halo baryon fraction and the entropy structure of the intracluster gas \citep{Rykoff08}. The $L_{X} - M$ relation has been extensively studied, mainly at low redshifts ($z \lesssim 0.1$) using X-ray data \citep{Markevitch98, Arnaud02,Reiprich02,Popesso05, Morandi07,Pratt09,Vikhlinin09}. The main conclusion was that the relation follows a power-law, but with a slope and amplitude that differ from the self-similar prediction of $M \propto L_{X}^{3/4}$. Instead, they found a flatter slope, $\alpha = 0.56 - 0.63$. Physical mechanism ruling the baryonic content of clusters, could strongly affect the X-ray luminosity, and so on the $L_{X} - M$ relation, causing deviations from a simple gravitational model. Simulations combining the gravitational evolution of dark matter structures together with the hydrodynamical behaviour \citep{Borgani04,Kay04,Borgani08} favor a lower slope value. 

Figure\,\ref{lumass} shows the $M-L_{X}$ relation for the galaxy clusters studied in this paper with masses estimated from the weak lensing analysis, together with those derived by other studies, the $M - L_{X}$ relation based on 12 low mass clusters from the CFHTLS by \citet{Kettula14}; 11 X-ray bright clusters selected and 206 stacked galaxy groups in the COSMOS field by \citet{Leauthaud10}, and the $L_{X} - M$ relation obtained from the EXCPRES sample by \citet{Foex12}. In principle, the slopes from $M - L_{X}$ ($\beta$) and $L_{X} - M$ ($\alpha$) could be easily compared ($\alpha = 1/\beta$) assuming that the halo mass function is locally a power-law \citep{Leauthaud10}. For comparison with other authors estimates, we used the NFW masses showed in Table\,\ref{table:2}. Given that \citet{Kettula14} derived core-excised luminosities, they are systematically lower than the rest of the plotted luminosities for a given mass. We notice that our mass determinations are in very good agreement with \citet{Leauthaud10} fit. The largest deviation from this fit corresponds to the two lowest X-ray luminosity clusters. Besides, [VMF98]093 contains a very low density of background galaxies which affects the precision of the shear estimates, and in the field of [VMF98]124 there is a star with X-ray emition, which could bias high the quoted X-ray luminosity of the cluster. 


\section{Summary and conclusions}
\label{sec:conclusions}

In this work we presented the weak lensing analysis of eight low X-ray luminosity galaxy cluster. We described the pipeline for determining weak lensing masses of clusters using ground-based images. The analysis consisted in: the detection and classification of the sources, the shape measurements on the $r'$ images taking into account the PSF, the galaxy background selection, the computation of shear profiles weighing the ellipticities according to the $r'$ magnitude and the colour of the galaxy, and finally, the fit of the mass density distribution models (SIS and NFW profiles). We have tested it succesfully on simulated data and then, we have applied it to a sample of low X-ray luminosity clusters. 

From this analysis we could estimate the mass of seven low X-ray luminosity galaxy clusters. One of these clusters ([VMF98]097) was previously analysed with a similar approach \citep{Carrasco07}. We improved the shear fit and we obtained a mass consistent with the previous result. For the other clusters in the sample, we estimated the mass for the first time.

The velocity dispersions obtained from the SIS fit, are in general agreement with the spectroscopic values available for four of the clusters in the sample. Masses obtained were compared to the X-ray luminosities. Our results are mostly in good agreement with previous analysis of the $M - L_{X}$ relation, in particular with \citet{Leauthaud10} result. In this work we provide further constraints for the $M - L_{X}$ relation, in low-intermediate X-ray luminosity galaxy clusters, by increassing the number of observables.

We plan in future works to include different models to fit the shear profiles, in order to include non-spherical models. Also, we plan to extend the pipeline to analyse low-massive galaxy systems employing stacking techniques.


\section*{Acknowledgments}
%
%
This work was partially supported by the Consejo Nacional de Investigaciones Cient\'{\i}ficas y T\'ecnicas (CONICET, Argentina) 
and the Secretar\'{\i}a de Ciencia y Tecnolog\'{\i}a de la Universidad Nacional de C\'ordoba (SeCyT-UNC, Argentina).\\
This research has made use of NASA's Astrophysics Data System and Cornell University arXiv repository.\\
Based on observations obtained at the Gemini Observatory processed using the Gemini IRAF package, which is operated by the Association of Universities for Research in Astronomy, Inc., under a cooperative agreement with the NSF on behalf of the Gemini partnership: the National Science Foundation (United States), the National Research Council (Canada), CONICYT (Chile), the Australian Research Council (Australia), Minist\'{e}rio da Ci\^{e}ncia, Tecnologia e Inova\c{c}\~{a}o (Brazil) and Ministerio de Ciencia, Tecnolog\'{i}a e Innovaci\'{o}n Productiva (Argentina). \\
We made an extensively use of the following python libraries:  http://www.numpy.org/, http://www.scipy.org/, http://roban.github.com/CosmoloPy/ and http://www.matplotlib.org/.\\
JLNC acknowledges the financial support from the Dirección de Investigación de la Universidad de La Serena (DIULS, ULS) and the Programa de Incentivo a la Investigación Académica (PIA-DIULS). Also acknowledges partial support from the postdoctoral fellow ALMA/CONICYT N 31120026.\\
Part of this research was developed while MJdLDR was postdoctoral researcher on the 
\href{Center for Gravitational Wave Astronomy at The University of Texas in Brownsville}{http://cgwa.phys.utb.edu/}.

\bibliographystyle{mn2e}
\bibliography{references}

\appendix

\end{document}